\documentclass[usenatbib]{mn2e}
\usepackage{graphicx,fixltx2e}

\def\rpd{\hbox{rad\,d$^{-1}$}}

\def\chisqr{\hbox{$\chi^2_{\rm r}$}}

\def\sn{\hbox{S/N}}
\def\vrad{\hbox{$v_{\rm rad}$}}
\def\ms{\hbox{m\,s$^{-1}$}}

\def\kms{\hbox{km\,s$^{-1}$}}
\def\vsini{\hbox{$v \sin i$}}

\def\ptt{\hbox{$10^{-4} I_{\rm c}$}}

\def\degr{\hbox{$^\circ$}}

\def\ts{\hbox{$t_{\rm s}$}}

\def\omeq{\hbox{$\Omega_{\rm eq}$}}
\def\dom{\hbox{$d\Omega$}}

\newcommand{\caii}{\hbox{Ca$\;${\sc ii}}}

\begin{document}

\title[Magnetic cycles of the planet-hosting star $\tau$~Bootis]
{Magnetic cycles of the planet-hosting star $\tau$~Bootis
\thanks{Based on observations 
obtained with ESPaDOnS at the Canada-France-Hawaii Telescope (CFHT) and with NARVAL at the 
T\'elescope Bernard Lyot (TBL).  CFHT/ESPaDOnS are operated by the National Research Council 
of Canada, the Institut National des Sciences de l'Univers of the Centre National de la 
Recherche Scientifique (INSU/CNRS) of France, and the University of Hawaii, while TBL/NARVAL 
are operated by INSU/CNRS.} }

\makeatletter

\def\newauthor{%
  \end{author@tabular}\par
  \begin{author@tabular}[t]{@{}l@{}}}
\makeatother

\author[J.-F.~Donati et al.]{\vspace{1.7mm} 
J.-F.~Donati$^1$\thanks{E-mail: 
donati@ast.obs-mip.fr (J-FD);
claire.moutou@oamp.fr (CM); 
rim.fares@ast.obs-mip.fr (RF); 
david.bohlender@nrc-cnrc.gc.ca (DB); 
claude.catala@obspm.fr (CC); 
magali.deleuil@oamp.fr (MD); 
shkolnik@ifa.hawaii.edu (ES);,
acc4@st-andrews.ac.uk (ACC);
mmj@st-andrews.ac.uk (MMJ)
gordonwa@uvic.ca (GAHW)},
C.~Moutou$^2$, R.~Far\`es$^1$, D. Bohlender$^3$, C.~Catala$^4$, M.~Deleuil$^2$, \\ 
\vspace{1.7mm}
{\hspace{-1.5mm}\LARGE\rm
E. Shkolnik$^5$, A.C.~Cameron$^6$, M.M.~Jardine$^6$, G.A.H.~Walker$^7$}\\
$^1$ LATT--UMR 5572, CNRS \& Univ.\ P.~Sabatier, 14 Av.\ E.~Belin, F--31400 Toulouse, France \\
$^2$ LAM--UMR 6110, CNRS \& Univ.\ de Provence, Traverse du Siphon, F--13376 Marseille, France \\
$^3$ HIA/NRC, 5071 West Saanich Road, Victoria, BC V9E 2E7, Canada \\ 
$^4$ LESIA--UMR 8109, CNRS \& Univ.\ Paris VII, 5 Place Janssen, F--92195 Meudon Cedex, France \\
$^5$ NASA Astrobiology Institute, Univ.\ of Hawaii at Manoa, 2680 Woodlawn Drive, Honolulu, HI 96822, USA  \\ 
$^6$ School of Physics and Astronomy, Univ.\ of St~Andrews, St~Andrews, Scotland KY16 9SS, UK \\
$^7$ 1234 Hewlett Place, Victoria, BC V8S 497, Canada
}

\date{2007, MNRAS, submitted}
\maketitle

\begin{abstract}
We have obtained new spectropolarimetric observations of the planet-hosting star $\tau$~Bootis, 
using the ESPaDOnS and NARVAL spectropolarimeters at the Canada-France-Hawaii Telescope (CFHT) and 
T\'elescope Bernard-Lyot (TBL).  

With this data set, we are able to confirm the presence of a 
magnetic field at the surface of $\tau$~Boo and map its large-scale structure over the whole 
star.  
The large-scale magnetic field is found to be fairly complex, with a strength of up to 10~G;  
it features a dominant poloidal field and a small toroidal component, the poloidal component 
being significantly more complex than a dipole.  
The overall polarity of the magnetic field has reversed 
with respect to our previous observation (obtained a year before), strongly suggesting that 
$\tau$~Boo is undergoing magnetic cycles similar to those of the Sun.  This is the first 
time that a global magnetic polarity switch is observed in a star other than the Sun;  
given the unfrequent occurrence of such events in the Sun, we speculate that the magnetic 
cycle period of $\tau$~Boo is much shorter than that of the Sun.  

Our new data also allow us to confirm the presence of differential rotation, both from 
the shape of the line profiles and from the latitudinal shearing that the magnetic structure is 
undergoing.  The differential rotation surface shear that $\tau$~Boo experiences is found 
to be 6 to 10 times larger than that of the Sun, in good agreement with recent claims that 
differential rotation is strongest in stars with shallow convective zones.  We propose that 
the short magnetic cycle period is due to the strong level of differential rotation.   

With a rotation period of 3.0 and 3.9~d at the equator and pole respectively, $\tau$~Boo 
appears as the first planet-hosting star whose rotation (at intermediate latitudes) is 
synchronised with the orbital motion of its giant planet (period 3.3~d).  Assuming that this 
synchronisation is not coincidental, it suggests that the tidal effects induced by the 
giant planet can be strong enough to force the thin convective enveloppe (though not the 
whole star) into corotation.  

We also detect time dependent activity fluctuations on $\tau$~Boo, but cannot unambiguously 
determine whether they are intrinsic to the star or induced by the planet;  more observations of 
similar type are needed to determine the role of the close-in giant planet orbiting $\tau$~Boo 
on both the activity enhancements and the magnetic cycle of the host star.  
\end{abstract}


\section{Introduction}

Magnetic fields of stars hosting close-in giant planets have recently started to trigger 
a lot of interest from the astrophysical community.  Magnetic fields on the host 
star are indeed likely to play a direct role in the survival of close-in giant planets.  
By evacuating the central regions of their protoplanetary discs (within 0.1~AU from the 
star typically, \citealt[e.g.,][]{Romanova06}), magnetic fields of young protostars are 
generating ideal conditions for giant planets to stop their inward migration whenever 
they enter the central gap of the disc, leaving them at a location where many close-in 
giant planets (making up 20\% of all known exoplanets to date) are actually observed.  

Magnetic fields of stars hosting close-in giant planets are also expected to be a key point 
in the way such planets interact with their host stars, and numerous studies have been carried 
out recently to study the details of this interaction.  Observations suggest that stars hosting 
close-in giant planets exhibit enhanced activity correlating with the 
orbital phase of the planet rather than with the rotation phase of the star, arguing that 
this increased activity is induced by the presence of the planet \citep[e.g.,][]{shk03, shk05};  
recent observations even suggest that these star-planet interactions could be cyclic in nature 
and oscillate between `on' and `off' states \citep{shk08}.  
Theoretical studies \citep[e.g.,][]{cuntz00,mcivor06} propose that this interaction could either result 
from tidal effects (e.g., enhancing turbulence and hence local dynamo action and activity within 
the planet-induced tidal bulge) or from magnetospheric interaction (e.g., inducing reconnection 
events as the planet travels through the large magnetospheric loops of the host star).  

For such studies, $\tau$~Boo (HR~5185, HD~120136, F7V) is an interesting candidate.  
Its massive planet is orbiting 
at 0.049~AU in 3.31~d \citep[e.g.,][]{Leigh03}.  Clear Zeeman signatures have recently been 
detected on $\tau$~Boo with the ESPaDOnS spectropolarimeter on the 3.6m Canada-France-Hawaii 
Telescope (CFHT), demonstrating that a large-scale field of a few G is present at the surface 
\citep{catala07}. Despite its relatively short rotation period (of order 3~d, \citealt{henry00, 
catala07}), the activity of $\tau$~Boo is only moderate (as usual for mid F stars) and shows no 
strong modulation with 
either orbital or rotation phase \citep{shk05, shk08};  the near synchronisation between the 
star's rotation and the planet orbit may explain this lack of planet-induced activity.  
$\tau$~Boo nonetheless appears as a good laboratory for studying and modelling the magnetic 
fields of stars with close-in giant planets, and in particular those of F stars with shallow 
convective zones and low intrinsic activity on which the (presumably very small) planet-induced 
activity enhancements are easier to detect.  

We present in the paper a detailed modelling of the magnetic field and differential rotation 
at the surface of $\tau$~Boo, as a follow-up study from the initial discovery of 
\citet{catala07}.  Sec.~\ref{sec:obs} presents the observations, Sec.~\ref{sec:mod} 
details the magnetic and differential rotation modelling while Sec.~\ref{sec:dis} 
provides a discussion and future prospects about this work.

\section{Observations}
\label{sec:obs}

For these renewed observations, we used again the ESPaDOnS spectropolarimeter (Donati et al., 
in preparation) on CFHT;  we also collected a few additional polarised spectra with NARVAL on 
the 2m T\'elescope Bernard Lyot (TBL).  
Both instruments yield a spectral resolution of about 65,000.  Each spectrum consists of 
a sequence of 4 individual subexposures taken in different configurations of the polarimeter 
retarders, in order to perform a precise and achromatic circular polarisation analysis and 
supress all spurious signatures at first order 
\citep[][Donati et al., in preparation]{donati97}.  

Data were reduced with the dedicated automatic reduction tool Libre-ESpRIT installed at CFHT 
and TBL \citep[][Donati et al., in preparation]{donati97}, 
and changed into sets of Stokes $I$ and $V$ spectra.  
All spectra are automatically corrected from spectral shifts resulting from
instrumental effects (eg mechanical flexures, temperature or pressure variations) using
telluric lines as a reference.  Though not perfect, this procedure allows spectra to be secured 
with a radial velocity (RV) precision of about 15--20~\ms\ \citep[e.g.,][]{catala07, moutou07}.

A total of 32 spectra were collected in 2007 June and July in variable weather conditions, 
mostly with ESPaDOnS/CFHT.  
The complete log is given in Table~\ref{tab:log}.  All data are phased with the same 
orbital ephemeris as that of \citet{catala07}:    
\begin{equation}
T_0 = \mbox{HJD~}2,453,450.984 + 3.31245 E
\label{eq:eph}
\end{equation}
with phase 0.0 denoting the first conjunction (i.e., with the planet farthest from the observer).  
Circularly polarised spectra of stars with stable magnetic topologies (e.g., chemically peculiar 
stars and hotter equivalents, like $\tau$~Sco, \citealt{donati06}) were observed during both runs 
to check that the instrument behaves properly and yields nominal field strengths and polarities.  

\begin{table*}
\caption[]{Journal of observations.  Columns 1--6 sequentially list the UT date,
the instrument used, the heliocentric Julian date and UT time (both at mid-exposure),
the complete exposure time 
and the peak signal to noise ratio (per 2.6~\kms\ velocity bin) of each observation.
Column 7 lists the rms noise level (relative to the unpolarized continuum level
$I_{\rm c}$ and per 1.8~\kms\ velocity bin) in the circular polarization profile
produced by Least-Squares Deconvolution (LSD), while column~8 and 9 list the
orbital cycle (using the ephemeris given by Eq.~\ref{eq:eph}) and the radial 
velocity (RV) associated with each exposure.  }    
\begin{tabular}{ccccccccc}
\hline
Date & Instrument & HJD          & UT      & $t_{\rm exp}$ & \sn\  & $\sigma_{\rm LSD}$ & Cycle & \vrad\\
(2007) &   & (2,453,000+) & (h:m:s) & (s) &      &   (\ptt)  & (245+) & (\kms) \\
\hline
Jun~12 & NARVAL/TBL    & 1264.45160 & 22:46:05 & $4\times600$ & 1,000 &  0.37 & 0.5788 & --16.506 \\

Jun~19 & NARVAL/TBL    & 1271.42056 & 22:02:04 & $4\times600$ &   910 &  0.42 & 2.6827 & --16.708 \\

Jun~26 & ESPaDOnS/CFHT & 1277.74494 & 05:49:50 & $4\times300$ & 1,550 &  0.23 & 4.5920 & --16.589 \\   
Jun~26 & ESPaDOnS/CFHT & 1277.81018 & 07:23:47 & $4\times300$ & 1,730 &  0.18 & 4.6117 & --16.645 \\

Jun~27 & ESPaDOnS/CFHT & 1278.74598 & 05:51:27 & $4\times200$ & 1,700 &  0.21 & 4.8942 & --16.644 \\
Jun~27 & ESPaDOnS/CFHT & 1278.75739 & 06:07:53 & $4\times200$ & 1,670 &  0.21 & 4.8976 & --16.641 \\
Jun~27 & ESPaDOnS/CFHT & 1278.76861 & 06:24:02 & $4\times200$ & 1,710 &  0.21 & 4.9010 & --16.625 \\

Jun~28 & ESPaDOnS/CFHT & 1279.73666 & 05:38:08 & $4\times200$ & 1,620 &  0.22 & 5.1933 & --15.927 \\
Jun~28 & ESPaDOnS/CFHT & 1279.74787 & 05:54:17 & $4\times200$ & 1,660 &  0.22 & 5.1966 & --15.915 \\
Jun~28 & ESPaDOnS/CFHT & 1279.75913 & 06:10:30 & $4\times200$ & 1,730 &  0.21 & 5.2000 & --15.912 \\

Jun~30 & ESPaDOnS/CFHT & 1281.83298 & 07:57:04 & $4\times200$ & 1,470 &  0.25 & 5.8261 & --16.746 \\
Jun~30 & ESPaDOnS/CFHT & 1281.84420 & 08:13:13 & $4\times200$ & 1,290 &  0.28 & 5.8295 & --16.739 \\

Jul~01 & ESPaDOnS/CFHT & 1282.73612 & 05:37:41 & $4\times200$ & 1,640 &  0.21 & 6.0988 & --16.070 \\
Jul~01 & ESPaDOnS/CFHT & 1282.74732 & 05:53:49 & $4\times200$ & 1,640 &  0.21 & 6.1022 & --16.062 \\
Jul~01 & ESPaDOnS/CFHT & 1282.75863 & 06:10:06 & $4\times200$ & 1,720 &  0.21 & 6.1056 & --16.058 \\

Jul~02 & ESPaDOnS/CFHT & 1283.73836 & 05:41:01 & $4\times200$ &   740 &  0.44 & 6.4013 & --16.093 \\
Jul~02 & ESPaDOnS/CFHT & 1283.74973 & 05:57:24 & $4\times200$ &   580 &  0.53 & 6.4048 & --16.090 \\
Jul~02 & ESPaDOnS/CFHT & 1283.76769 & 06:23:16 & $4\times200$ & 1,240 &  0.31 & 6.4102 & --16.103 \\
Jul~02 & ESPaDOnS/CFHT & 1283.78209 & 06:43:60 & $4\times200$ & 1,360 &  0.27 & 6.4145 & --16.110 \\
Jul~02 & ESPaDOnS/CFHT & 1283.91960 & 10:02:02 & $4\times300$ & 1,370 &  0.29 & 6.4561 & --16.193 \\

Jul~03 & ESPaDOnS/CFHT & 1284.74158 & 05:45:46 & $4\times300$ & 1,370 &  0.24 & 6.7042 & --16.816 \\
Jul~03 & ESPaDOnS/CFHT & 1284.75763 & 06:08:53 & $4\times300$ & 1,830 &  0.19 & 6.7090 & --16.818 \\
Jul~03 & ESPaDOnS/CFHT & 1284.77363 & 06:31:56 & $4\times300$ & 1,750 &  0.21 & 6.7139 & --16.808 \\
Jul~03 & ESPaDOnS/CFHT & 1284.85319 & 08:26:30 & $4\times300$ & 1,700 &  0.23 & 6.7379 & --16.817 \\

Jul~04 & ESPaDOnS/CFHT & 1285.79220 & 06:58:47 & $4\times300$ & 1,090 &  0.33 & 7.0214 & --16.275 \\
Jul~04 & ESPaDOnS/CFHT & 1285.80803 & 07:21:35 & $4\times300$ & 1,470 &  0.26 & 7.0262 & --16.248 \\
Jul~04 & ESPaDOnS/CFHT & 1285.86573 & 08:44:40 & $4\times300$ & 1,240 &  0.32 & 7.0436 & --16.191 \\
Jul~04 & ESPaDOnS/CFHT & 1285.88171 & 09:07:41 & $4\times300$ &   990 &  0.41 & 7.0484 & --16.173 \\

Jul~05 & ESPaDOnS/CFHT & 1286.74025 & 05:44:05 & $4\times200$ &   330 &  1.18 & 7.3076 & --15.943 \\
Jul~05 & ESPaDOnS/CFHT & 1286.75144 & 06:00:12 & $4\times200$ &   250 &  1.57 & 7.3110 & --15.933 \\
Jul~05 & ESPaDOnS/CFHT & 1286.76265 & 06:16:20 & $4\times200$ &   310 &  1.28 & 7.3143 & --15.930 \\
Jul~05 & ESPaDOnS/CFHT & 1286.78415 & 06:47:18 & $4\times600$ &   400 &  0.98 & 7.3208 & --15.941 \\
\hline
\end{tabular}
\label{tab:log}
\end{table*}

Least-Squares Deconvolution \citep[LSD, ][]{donati97} was applied to all spectra to 
retrieve average unpolarised and circularly polarised profiles of photospheric spectral 
lines.  The line pattern used for this process is derived from a Kurucz model atmosphere 
with solar abundances, and effective temperature and logarithmic gravity set to 6250~K 
and 4.0 respectively;  this line pattern includes most moderate to strong lines present 
in the optical domain (those featuring central depths larger than 40\% of the local 
continuum, before any macroturbulent or rotational broadening, about 4,000 throughout 
the whole spectral range) but excludes the very strongest, broadest features, such as 
Balmer lines, whose Zeeman signature is strongly smeared out compared to those of 
narrow lines. The typical multiplex gain in S/N for polarisation profiles is about 25 to 
30, implying noise levels in LSD polarisation profiles as low as 20~ppm (i.e., $2\times10^{-5}$ 
in units of the unpolarised continuum) in good weather conditions.  Zeeman signatures, 
whenever detected, have a typical amplitude of about 100~ppm (see below).

Radial velocities are obtained from Gaussian fits to each LSD unpolarised profile of 
$\tau$~Boo.  We find that our measurements are in good agreement with previous measurements
(see Fig.~\ref{fig:vrad}).  Residuals (rms) with respect to the predicted RV curve are equal to 20~\ms, 
i.e., similar to the uncertainties of 15--20~\ms\ reported by \citet{catala07} and \citet{moutou07}.  
All Stokes $I$ and $V$ profiles used in the following 
were corrected for the orbital motion, i.e., translated into the velocity rest frame of 
$\tau$~Boo.  

\begin{figure*}
\center{\hbox{\includegraphics[scale=0.37,angle=-90]{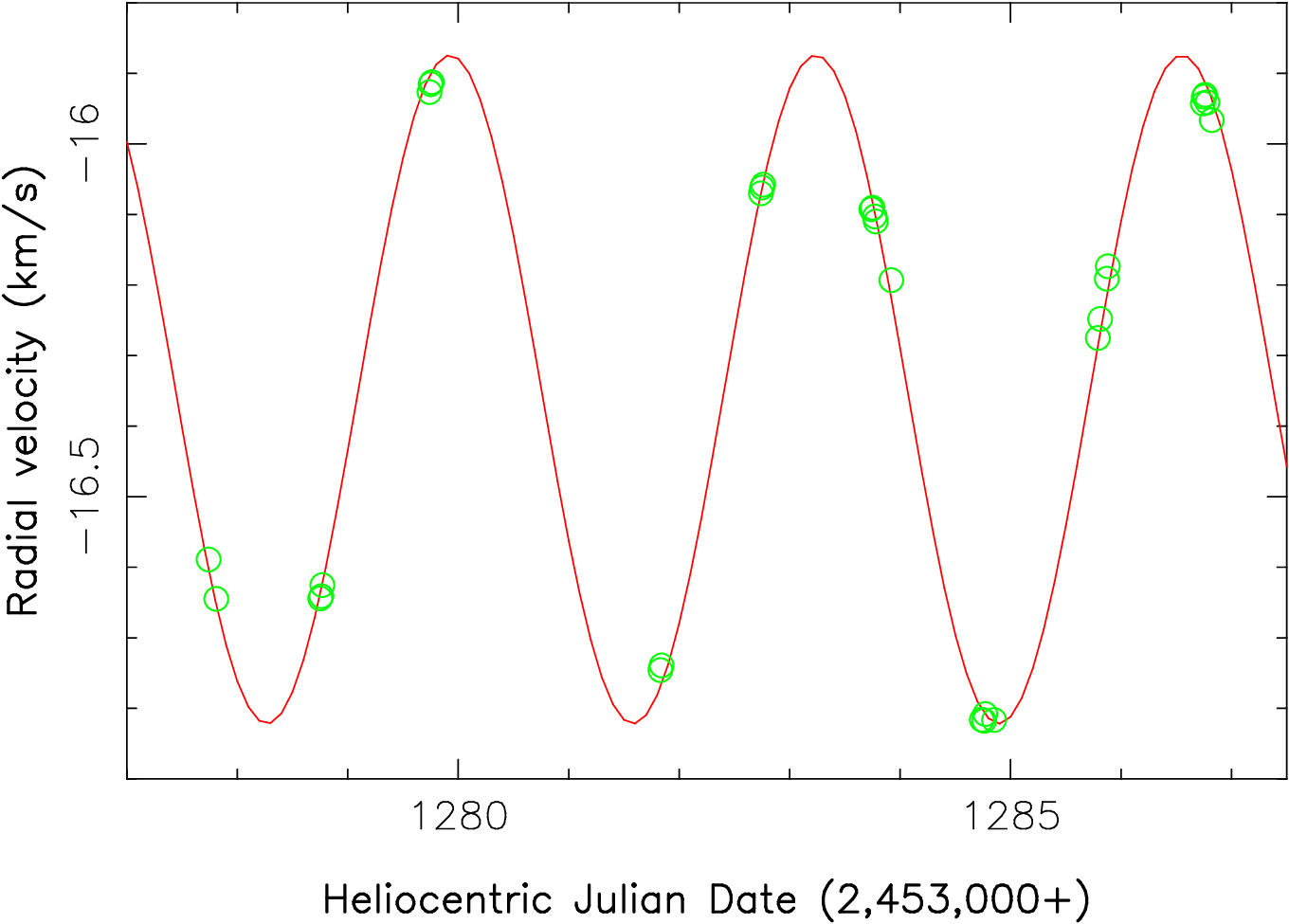}\hspace{3mm}
              \includegraphics[scale=0.37,angle=-90]{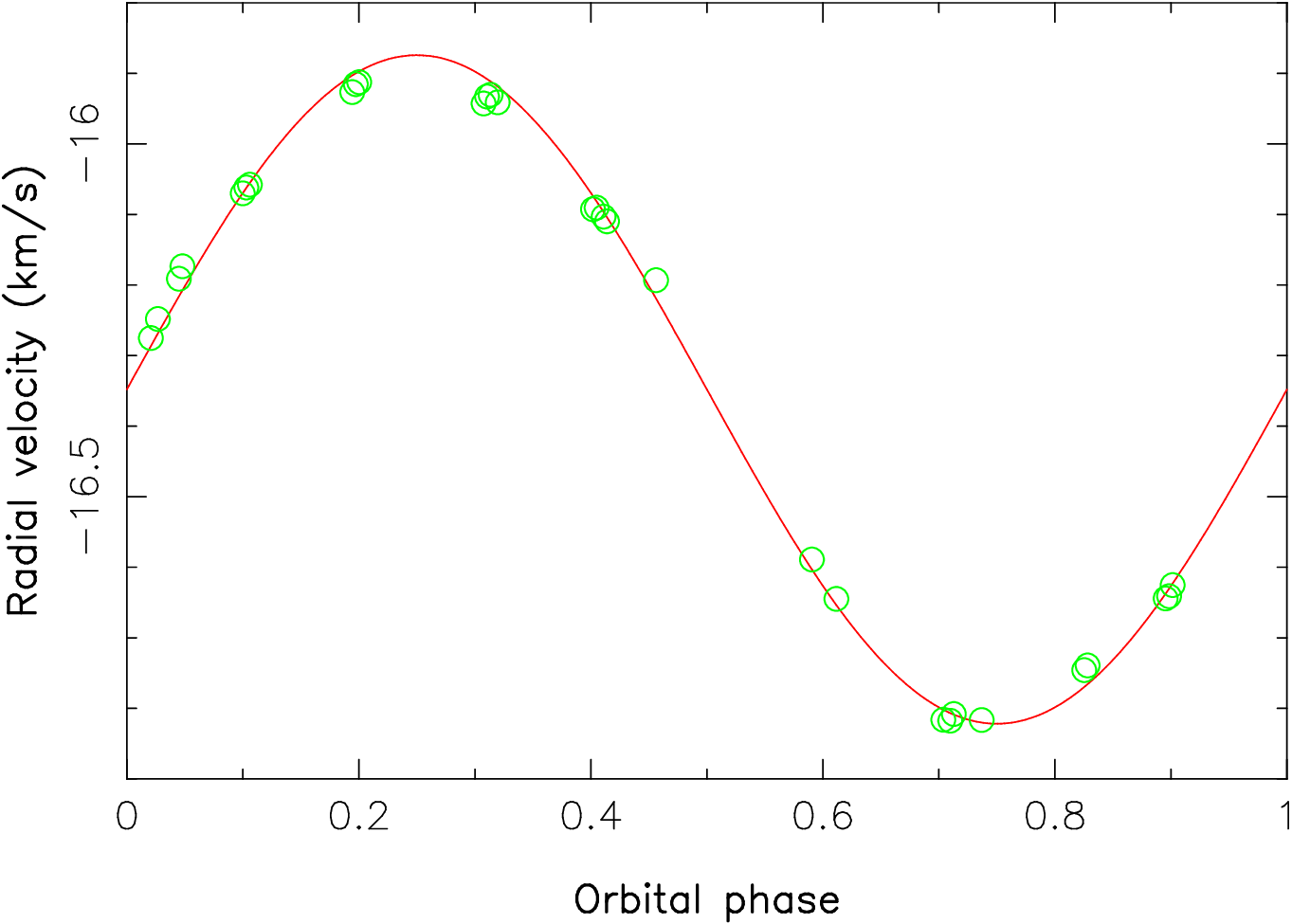}}} 
\caption[]{Radial velocities of $\tau$~Boo derived from our ESPaDOnS spectra 
(green circles) as a function of Heliocentric Julian Date (left) and orbital phase 
(right, using the ephemeris of eq.~\ref{eq:eph}).  The radial velocity model  
plotted here (red full line) corresponds to a circular orbit with an (assumed) 
velocity semi-amplitude of 467~\ms\ \citep{butler97} and a (fitted) systemic 
velocity of $-16.348$~\kms.  Individual 
error bars on data points (15--20~\ms) are about as large as the symbol size. }
\label{fig:vrad}
\end{figure*}

As in \citet{catala07}, we assume that $\tau$~Boo rotates about 
an axis inclined at an angle $i=40$\degr\ with respect to the 
line-of-sight.  The unpolarised spectral lines are significantly broadened by rotation 
and suggest that the rotation of the star is more or less synchronised with the planet 
orbital motion, i.e., corresponds to a rotation rate of 1.9~\rpd.  
By averaging all Stokes $I$ LSD profiles into a single mean line and computing its 
Fourier transform, one can estimate 
the amount of differential rotation at the surface of the star \citep{reiners02}.  
This experiment, first carried out by \citet{reiners06} and repeated by \citet{catala07}, 
demonstrates that $\tau$~Boo is indeed differentially rotating, with a relative differential 
rotation (i.e., an angular rotation shear relative to the mean angular rotation rate) 
of 18--20\% and an angular rotation shear\footnote{The angular rotation 
shear derived by \citet{reiners06}, equal to $0.31\pm0.13$~\rpd, 
is actually scaled down by $\sqrt{\sin i}$;  using $i=40$\degr, 
it translates into a true angular rotation shear of $0.38\pm0.16$~\rpd\, 
in good agreement with our own estimate.} of $0.35\pm0.10$~\rpd;  the corresponding rotation 
periods at the equator and the pole are equal to about 3.0~d and 3.7~d respectively, 
bracketing the planet orbital period of 3.3~d.  The Fourier transform of the average 
LSD Stokes $I$ profile derived from our new data (not shown here) is very similar to that 
shown in \citet{catala07} (their Fig.~1) and confirms their analysis.

\section{Magnetic modelling}
\label{sec:mod}

\subsection{Model description}

To model the magnetic topology of $\tau$~Boo, we use the new imaging code of \citet{donati06} 
where the field topology is described through spherical-harmonics expansions.  We use the 
principles of maximum entropy image reconstruction to retrieve the simplest magnetic image 
compatible with the series of rotationally modulated Zeeman signatures.  
More specifically, the field is divided into its radial-poloidal, non-radial-poloidal and 
toroidal components, each of them described as a spherical-harmonics expansion;  given 
the different rotational modulation of Zeeman signatures that poloidal and toroidal fields 
generate, our imaging code appears particularly useful and efficient at producing 
dynamo-relevant diagnostics about the large-scale magnetic topologies at the surface of 
late-type stars.  One of the latest application of this code can be found in \citet{morin07}.  

The reconstruction process is iterative and proceeds by comparing at each step the synthetic 
profiles corresponding to the current image with those of the observed data set.  To compute 
the synthetic Zeeman signatures, we divide the surface of the star into small grid cells 
(typically a few thousands), work out the specific contribution of each grid cell to the Stokes 
$V$ profiles (given the magnetic field strength and orientation within each grid cell, as well 
as the cell radial velocity, location and projected area) and finally sum up contributions of 
all cells.  Since the problem is partly ill-posed, we stabilise the inversion process by using 
an entropy criterion (applied to the spherical harmonics coefficients) aimed at selecting the 
image with minimum information among all those compatible with the data \citep{morin07}.  

The model we use to describe the local Stokes $I$ and $V$ profiles is quite simple.  While the 
local unpolarised profile is given by a Gaussian, the local circular polarisation profile is 
computed assuming the weak field approximation, i.e., that $V$ is proportional to 
$dI/d\lambda$ and to the local line-of-sight component of the magnetic field 
\citep[e.g.,][]{donati97}.  Assuming a line-of-sight projected equatorial rotation velocity 
\vsini\ of 15.9~\kms\ and a relative differential rotation of 18\%, we obtain synthetic Stokes 
$I$ profiles whose first 2 zeros in the Fourier transform match those of the observed data;  
setting the full width at half maximum of the local Stokes $I$ profile to 11~\kms\ produces a 
very good fit to the average Stokes $I$ LSD profiles.  This simple line model was used quite 
extensively and has proved to be efficient at correctly reproducing observed sets of 
rotationally modulated  Zeeman signatures \citep[e.g.,][]{catala07, moutou07}.  

To incorporate differential rotation into the modelling, we proceed as in \citet{donati03} and 
\citet{morin07}, i.e., 
by assuming that the rotation rate at the surface of the star is varying with latitude 
$\theta$ as  $\omeq - \dom \sin^2 \theta$ where \omeq\ is the rotation rate at the equator 
\dom\ the difference in rotation rate between the equator and the pole.  When computing the 
synthetic profiles, we use this law to work out the longitude shift of each cell at each 
observing epoch with respect to its location at the median observing epoch (at which the field 
is reconstructed, i.e., orbital cycle $6+245=251$ in the present case and in the ephemeris of 
eq.~\ref{eq:eph} or HJD~2,454,282.41) 
so that we can correctly evaluate the true spectral contributions of all cells at all epochs.  

For each pair of \omeq\ and \dom\ values within a range of acceptable values, we then derive, 
from the complete data set, the corresponding magnetic topology (at a given information content) 
and the associated reduced chisquare level \chisqr\ at which modelled spectra fit observations.  
By fitting a paraboloid to the \chisqr\ surface derived in this process \citep{donati03}, we can 
easily infer the magnetic topology that yields the best fit to the data along with the corresponding
differential rotation parameters and error bars.  This process has proved fairly reliable to 
estimate surface differential rotation on magnetic stars \citep[e.g.,][]{donati03}.

\subsection{Results}

We applied this model to our complete set of Zeeman signatures of $\tau$~Boo.  Data obtained 
on the last night (Jul~05), i.e., in bad weather conditions, were finally excluded from the fit 
as they provide very little information (given their low quality with respect to the bulk of 
the data set, see Table~\ref{tab:log}).  The rotational broadening of spectral lines provides 
significant spatial resolution at the surface of the star, i.e., from 9 to 22 resolution 
elements around the equator depending on whether we use the intrinsic line profile 
(11~\kms) or the instrumental profile (4.5~\kms) to define the resolution element.  
The spherical harmonics expansions used to describe the magnetic 
field were first computed up to orders $\ell=15$;  in practice, little improvement is 
obtained when adding terms with orders higher than 8 in the expansion.  All results presented 
below were thus derived with expansions limited to $\ell=8$.  

The first result from this modelling is that the Stokes $V$ data provide a completely independent 
confirmation of both the existence and magnitude of differential rotation at the surface of the 
star, as demonstrated by the well defined \chisqr\ paraboloid we obtain (see Fig.~\ref{fig:drot}).  
The differential rotation parameters producing an optimal fit to the data (at a given information 
content in the magnetic map) are respectively equal to $\omeq=2.10\pm0.04$~\rpd\ and 
$\dom=0.50\pm0.12$~\rpd\ and imply a relative differential rotation of 24\%.  The corresponding 
rotation periods at the equator and pole are equal to about 3~d and 3.9~d respectively;  
the optimal period assuming solid body rotation is 3.23~d and corresponds to the average recurrence 
time of the detected Zeeman signatures, i.e.\ to the rotation period at the average latitude 
(about 35\degr) of the reconstructed magnetic structures  \citep{donati03}.  
The latitudinal surface shear we derive is about 8 to 10 times stronger than that of the Sun;  
the corresponding timescale for the equator to lap the pole by one complete rotation cycle is 
only about 12~d.  

\begin{figure}
\center{\includegraphics[scale=0.35,angle=-90]{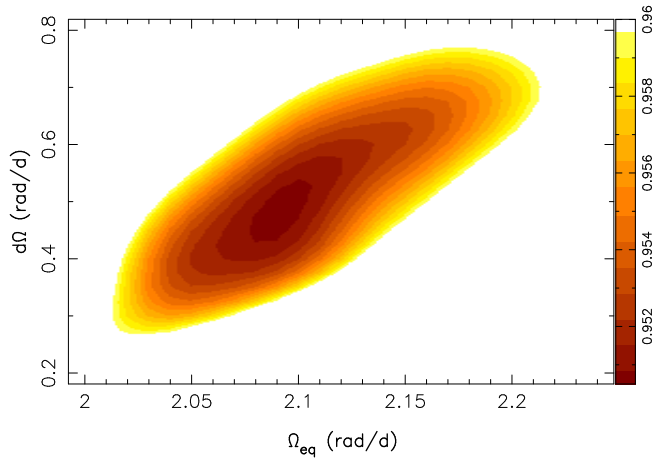}}
\caption[]{Variation of \chisqr\ as a function of \omeq\ and \dom, derived from the modelling 
of our Stokes $V$ data set on $\tau$~Boo.  The outer colour contour corresponds to a 1\% increase 
in \chisqr, and trace a 2.7$\sigma$ interval for each parameter taken separately, or a 1.7$\sigma$ 
interval for both parameters as a pair.   }  
\label{fig:drot}
\end{figure}

\begin{figure*}
\center{\includegraphics[scale=0.7]{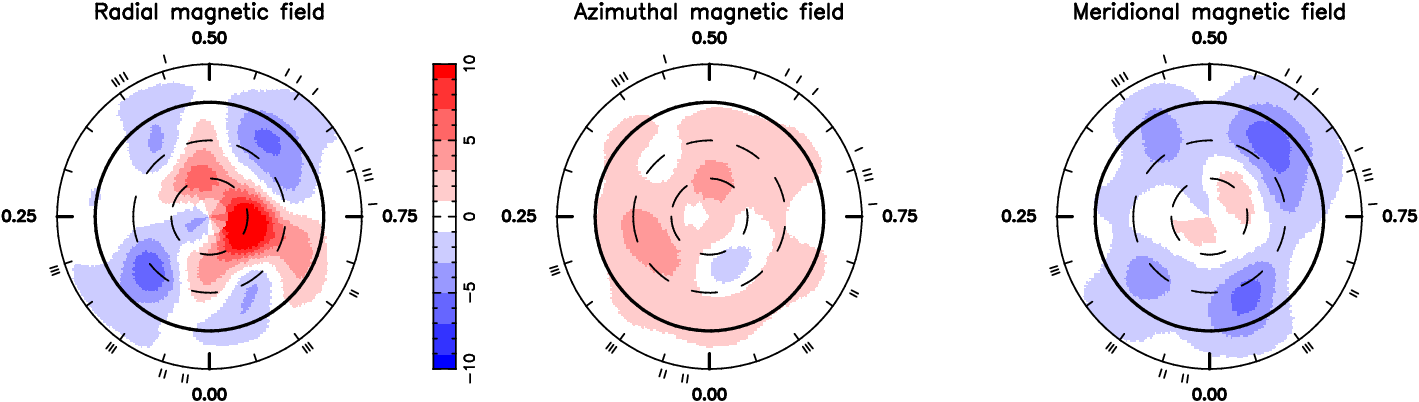}}
\caption[]{Maximum-entropy reconstructions of the large-scale magnetic topology of
$\tau$~Boo as derived from our 2007 data set (at orbital cycle 6+245 or HJD~2,454,282.41).  
The radial, azimuthal and meridional components of the field are displayed from 
left to right (with magnetic flux values labelled in G).
The star is shown in flattened polar projection down to latitudes
of $-30\degr$, with the equator depicted as a bold circle and
parallels as dashed circles.  Radial ticks around each plot indicate
orbital phases of observations.  } 
\label{fig:map}
\end{figure*}

\begin{figure*} 
\center{\includegraphics[scale=0.7,angle=-90]{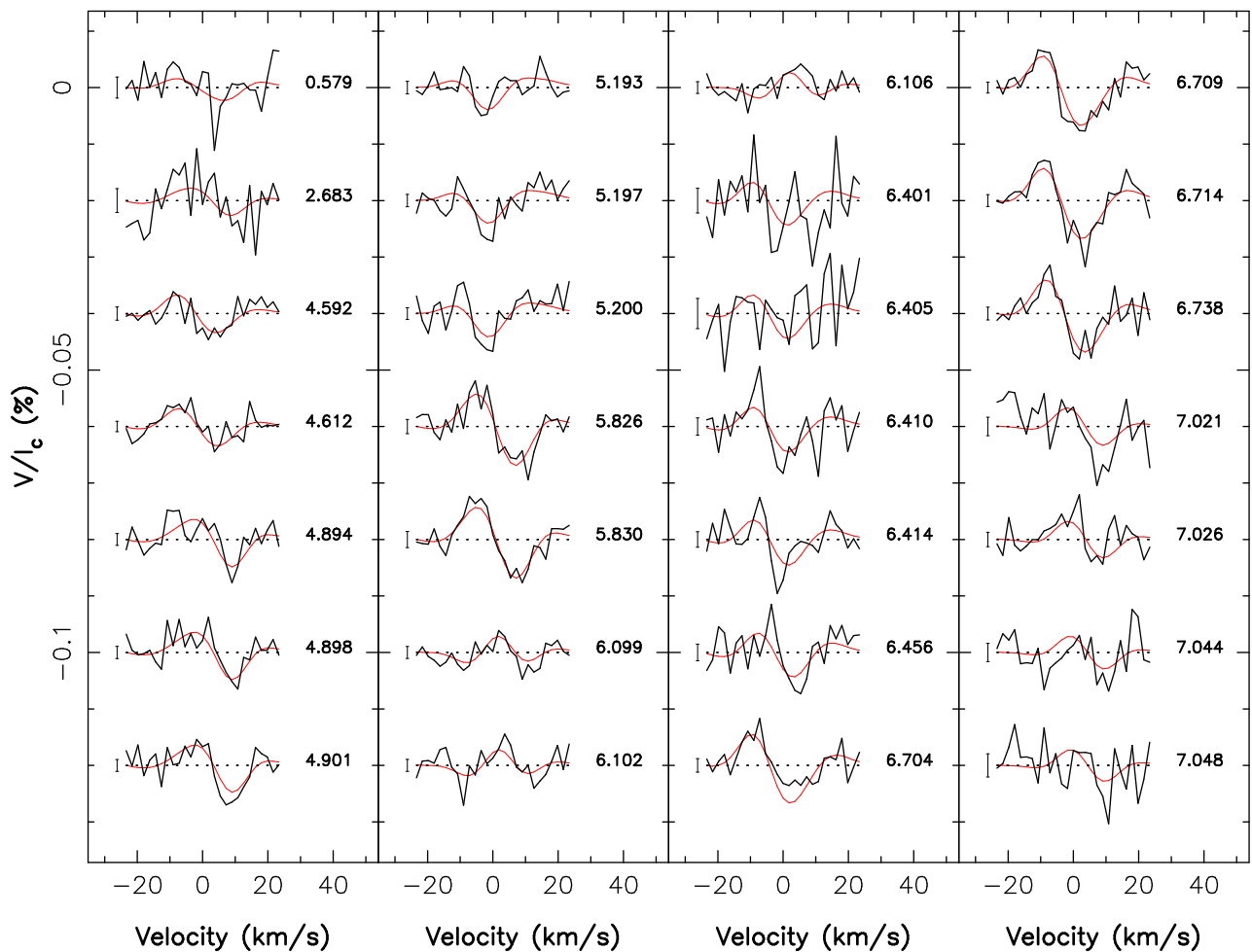}}
\caption[]{Maximum-entropy fit (thin red line) to the observed Stokes $V$ LSD 
profiles (thick black line) of $\tau$~Boo.  The orbital cycle of each observation 
(as listed in Table~1) and 1$\sigma$ error bars are also shown next to each profile.  }
\label{fig:fit}
\end{figure*}

The large-scale magnetic map we derive (see Fig.~\ref{fig:map}) corresponds to a unit \chisqr\ fit 
to the data (see Fig.~\ref{fig:fit}), the initial \chisqr\ being about 1.9 for 750 data points.  
Note that the evolution of the magnetic field under the 
shearing effect of differential rotation throughout the period of our observations can be traced 
directly to the data themselves;  for instance, 
the Zeeman signature collected at cycle 2.68+245 is significantly 
smaller in amplitude than those collected 4 rotation cycles later (around cycles 6.70+245).  

We find that the field topology includes a small toroidal component 
enclosing 17\% of the overall reconstructed magnetic energy;  
while the Stokes $V$ data can be fitted without the toroidal component, the corresponding 
reconstructed map contains significantly more (i.e., $+50$\%) information than that of 
Fig.~\ref{fig:map}, suggesting that a purely poloidal field is less probable and that the 
reconstructed toroidal component is likely real.  The toroidal component is clearly visible 
in Fig.~\ref{fig:map} and shows up as ring of positive (i.e., counterclockwise) azimuthal 
field encircling the star at mid latitude.  

The magnetic field is obviously more complex than a dipole, $\ell=1$ modes enclosing only about 
30\% of the poloidal field energy.  The quadrupolar and octupolar terms (i.e.\ $\ell=2$ and 
$\ell=3$ modes) dominate the field distribution in the visible hemisphere and contain 40\% of 
the reconstructed poloidal field energy, while the remaining 30\% spreads into higher order 
terms.  This is directly visible from Fig.~\ref{fig:map};  the radial field map features 
a main positive pole at high latitudes surrounded by an incomplete ring of negative field at 
low latitudes, reminiscent of a slightly tilted quadrupole or octupole.  
We also find that the reconstructed poloidal field is mostly axisymmetric with 
respect to the rotation axis;  modes verifying  $m<\ell/2$ are enclosing a dominant fraction 
(i.e.\ 60\%) of the poloidal field energy, while the non axisymmetric modes (with $m>\ell/2$) 
contain no more than 30\% of the poloidal field energy.  This is again fairly obvious from 
the reconstructed map of Fig.~\ref{fig:map}.

\section{Discussion and conclusion}
\label{sec:dis}

Thanks to this new data set, we achieved a number of significant results relevant to 
dynamo processes and magnetic field generation in cool stars with shallow convective zones;  
We summarise them below and discuss their implications for the study of star/planet 
magnetic interactions in systems hosting close-in giant planets.  

First, we obtained 2 completely independent estimates of the differential 
rotation at the surface of $\tau$~Boo.  From the temporal distortion of the large-scale magnetic 
topology, we find that the latitudinal angular rotation shear is equal to $\dom=0.50\pm0.12$~\rpd, 
i.e., about 8 to 10 times that of the Sun.  This is in reasonable  agreement with the estimate 
derived from the detailed shape of spectral lines and of their Fourier transform, yielding 
$\dom=0.35\pm0.10$~\rpd.  It unambiguously demonstrates that $\tau$~Boo is experiencing strong 
differential rotation at photospheric level;  this is apparently a general trend of early G and 
F stars \citep[e.g.,][]{marsden06, reiners06}.  Our measurement is also the first direct 
and simultaneous confirmation that both methods employed up to now to investigate latitudinal 
shears on stellar surfaces are actually yielding consistent results.  
The rotation period at the equator of $\tau$~Boo is 3~d, i.e., about 10\% shorter than the 
orbital period of the giant planet.  It confirms in particular the estimate first obtained by 
\citet{catala07};  it also implies that the giant planet is synchronised with the surface 
of the star at a latitude of about 40\degr.  

We also derived how the magnetic field is distributed at the surface of $\tau$~Boo.  In particular, 
we find that the magnetic field is mostly poloidal despite the vigorous differential rotation.  
This is different than what \citet{marsden06} report for another cool star with a shallow 
convective zone (HD~171488), in which the magnetic field is apparently distributed roughly evenly 
between poloidal and toroidal field components;  the main difference between both stars is the 
rotation rate, less than half as large for $\tau$~Boo than for HD~171488.  
The magnetic topology we reconstruct is grossly similar to (though much more accurate than) 
that derived by \citet{catala07} from a much sparser data set.  There is however one major 
difference between both maps;  while the radial field is predominantly positive and negative 
at high and low latitudes respectively in our image, the opposite holds in the 2006 
image reconstructed by \citet{catala07}.  A similar polarity inversion is observed for the 
two other field components between the 2006 and 2007 images.  Check stars with known magnetic 
polarities observed during both runs demonstrate that this global polarity switch is not 
due to instrumental or data reduction problems and can only be attributed to $\tau$~Boo itself.  

This is the first time that a global magnetic polarity switch is observed in a star other than 
the Sun.  Given that such events are rather unfrequent in the Sun (only once every 11~yr) and 
have never been detected yet in the 20 or so stars (of various spectral types) 
that have been observed more than once up to now \citep[e.g.,][]{donati03b}, 
we speculate that the magnetic cycle of $\tau$~Boo is likely shorter than that of the Sun.  
If the cycle period varies more or less linearly with the strength of differential rotation, 
we expect the period of the full magnetic cycle (22~yr in the case of the Sun) to be of order 
2--3~yr in $\tau$~Boo;  this argues for renewed and regular 
spectropolarimetric observations of $\tau$~Boo to monitor the evolution of the magnetic field 
throughout a complete magnetic cycle.  Given that the convective zone of $\tau$~Boo is very 
shallow and essentially reduces to a thin layer similar in nature to the solar tachocline 
(where the magnetic dynamo is expected to operate mostly), our result directly demonstrates 
that interface dynamos are indeed capable of producing oscillating magnetic topologies.  

At this stage, there is not much we can say about activity putatively induced by the 
presence of the giant planet.  There is actually no low-latitude magnetic features either facing 
the planet (at phase 0.5) or on the other side of the star (phase 0.0), i.e., in regions at 
which tidal effects are maximum.  The main low-latitude magnetic features we detect are the 
negative radial and meridional field features at phases 0.13, 0.40, 0.60 and 0.93 
(see Fig.~\ref{fig:map});  however, 
these features are apparently rotating faster than the planet orbital motion (being those 
from which differential rotation at low latitudes is estimated) and 
can therefore not be interpreted as due to a putative tidal bulge (rotating 
in phase with the orbital motion).  

We find that activity signatures in usual spectral indexes (H$\alpha$, 
\caii\ H \& K, and infrared triplet lines) are very weak, smaller than 0.5\% of the unpolarised 
continuum;  they are best visible in H$\alpha$ thanks to the higher spectrum quality around 
700~nm (see Fig.~\ref{fig:hal}) and reach a peak to peak amplitude of about 0.2~\kms\ 
(0.44~pm) only.  Maximum H$\alpha$ emission (i.e.\ positive integrals over the residual 
spectra of Fig.~\ref{fig:hal}) occurs roughly twice per rotation 
(at phases 0.1 and 0.7) and roughly coincide with the main low-latitude radial 
field features seen in the magnetic map;  we therefore suspect that the two small activity 
enhancements we detect trace intrinsic activity from the star itself.  It is unlikely 
that this activity is induced by the planet itself through tidal friction (also 
expected to produce two activity enhancements around phase 0.0 and 0.5);  given that the 
stellar equator is rotating faster than the orbital motion, the tidal bulge is expected to 
be slightly ahead of the planet (shifted to negative orbital phases), i.e., the opposite of 
what we actually see.  The near synchronisation between the planet orbital motion 
and the star equatorial rotation (beat period of about 32~d) likely implies that the tidal 
bulge generates very little friction and features only a marginal misalignment with the 
star-planet direction.  Magnetic reconnection triggered by the planet nevertheless remains 
a potential option for explaining the observed activity variations.    

\begin{figure}
\center{\includegraphics[scale=0.75,angle=-90]{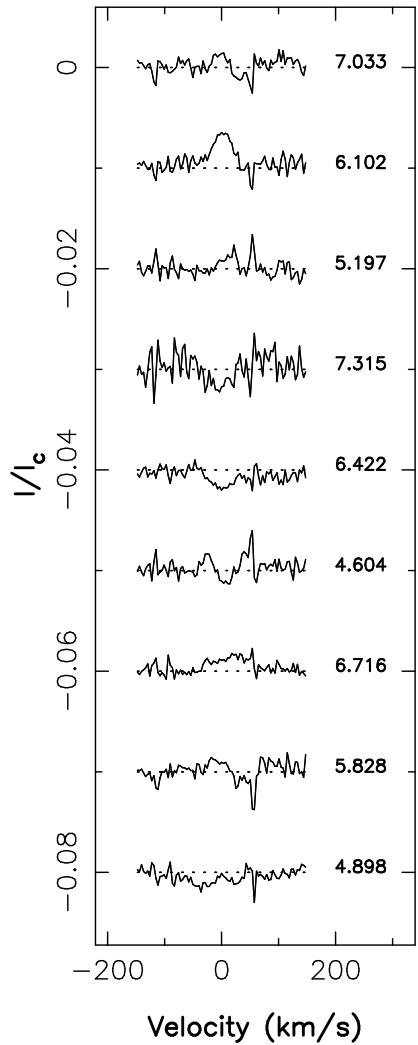}}
\caption[]{Residual H$\alpha$ signature as a function of orbital cycle.  Profiles recorded 
on the same night were averaged together to reduce noise, and a mean H$\alpha$ 
profile was subtracted from each nightly average.  This operation was performed on ESPaDOnS 
spectra only. The stray pixels at $+50$~\kms\ are due to a weak (and variable) telluric line.  } 
\label{fig:hal}
\end{figure}

We note that $\tau$~Boo is apparently the first convincing case of a star whose rotation 
is synchronised with the orbital motion of its close-in giant planet.  
Following \citet{zahn94} and \citet{marcy97}, the synchronisation timescales \ts\ (in yr) 
it takes for the planet to enforce corotation of the whole star can be approximated by:  
\begin{equation}
\ts=4.5\left( \frac{M}{m}\right) ^2 \left(\frac{a}{R}\right) ^6 
\end{equation}
where $M/m$ is the star to planet mass ratio (equal to about $350 \sin i$ for $\tau$~Boo) and $a/R$ 
the planet orbit semi-major axis relative to the radius of the star (equal to about 7.2 for 
$\tau$~Boo).  Using $i=40\degr$, we obtain that $\ts\simeq30$~Gyr, i.e., much larger than 
the estimated lifetime of $\tau$~Boo (about 1~Gyr).  Since we consider that the apparent 
spin-orbit synchronisation between $\tau$~Boo and its close-in giant planet is unlikely 
to be coincidental, we suspect the synchronisation timescale to be likely overestimated;  
one possible reason is that synchronisation is not achieved on the whole star (as assumed 
by \citealt{zahn94}) but only on a restricted volume close to the surface of the star, e.g., 
the convective zone proper whose mass is estimated to less than 0.1\% that of the star.  
If the tidal effects induced by the close-in giant planet on its host star are strong enough 
to trigger synchronisation of at least the shallow convective zone, they may also play a 
significant role in the dynamo processes operating in this thin layer.  
  
More high-quality observations densely sampling the orbital and rotation cycles, carried 
out over a timescale of typically a month, and repeated at least once a year, are required to 
go further along these tracks.  Such data sets will first allow us to obtain a complete 
magnetic monitoring of the activity cycle of $\tau$~Boo and give us the opportunity to 
achieve the first such study on a star other than the Sun.  These data should also enable 
us to estimate the lifetime and recurrence rate of the activity enhancements that we 
detected on $\tau$~Boo;  if these episodes turn out to be long-lived and synchronised with 
the orbital motion (rather than with the rotation rate at the stellar equator), 
they could be unambiguously attributed to the giant planet.  Finally, carrying out similar 
observations on a sample of stars with and without planets to look for statistical differences 
between both subsamples will also be necessary to investigate in more details the impact of 
close-in giant planets on the dynamo processes of stars with shallow convective zones.  

\section*{Acknowledgments}

We thank N.~Letourneur and J.-P.~Michel for collecting the NARVAL data for us, and the CFHT and 
TBL staff for their help during the observations.  We are grateful to the referee, 
A.~Lanza, for providing comments that improved the manuscript.  

\bibliography{tbnew}
\bibliographystyle{mn2e}

\end{document}